\documentclass[aps,pra,twocolumn,amsmath,amssymb,showpacs,10pt]{revtex4-1}
\usepackage{times}

\usepackage[usenames,dvipsnames]{xcolor}
\newcommand{\Brev}[1]{\textcolor{black}{#1}}
\newcommand{\BVrev}[1]{\textcolor{black}{#1}}
\newcommand{\CUrev}[1]{\textcolor{black}{#1}}
\newcommand{\GGrev}[1]{\textcolor{black}{#1}}

\usepackage[english]{babel}
\usepackage{graphicx}
\usepackage{dcolumn}
\usepackage{bm}
\usepackage{color}
\usepackage{amsmath}
\usepackage{relsize,amsmath,mathrsfs,empheq,verbatim}
\usepackage{etoolbox}
\usepackage[caption = false]{subfig}
\apptocmd{\thebibliography}{\raggedright}{}{}

\newcommand{\bra}[1]{\langle #1 |}
\newcommand{\ket}[1]{| #1 \rangle}

\newcommand{\mean}[1]{\langle #1 \rangle}
\providecommand{\bgreek}[1]{\mbox{\boldmath$#1$}}

\newcommand{\Ham}{\mathcal{H}}

\begin{document}
\newdimen\origiwspc%
  \newdimen\origiwstr%

\title{Energy backflow and non-Markovian dynamics}

\author{G. Guarnieri$^{1,2}$, C. Uchiyama $^{3}$, B. Vacchini$^{1,2}$}

\affiliation{$^1$\mbox{Dipartimento di Fisica, Universit{\`a} degli Studi di Milano, Via Celoria 16, 20133 Milan, Italy}\\
$^2$\mbox{Istituto Nazionale di Fisica Nucleare, Sezione di Milano, Via Celoria 16, 20133 Milan, Italy}\\
$^3$\mbox{Graduate School of Interdisciplinary Research,}\\
\mbox{University of Yamanashi, 4-3-11, Takeda, Kofu, Yamanashi 400-8511, Japan}}

\begin{abstract} We explore the behavior in time of the energy exchange between a system of interest and its environment, together with its relationship to the non-Markovianity of the system dynamics.  In order to evaluate the energy exchange we rely on the full counting statistics formalism, which we use to evaluate the first moment of its probability distribution.  We focus in particular on the energy backflow from environment to system, to which we associate a suitable condition and quantifier, which enables us to draw a connection with a recently introduced notion of non-Markovianity based on information backflow. This quantifier is then studied in detail in the case of the spin-boson model, described within a second order time-convolutionless approximation, \GGrev{observing that non-Markovianity allows for the observation of energy backflow.} This analysis allows us to identify the parameters region in which energy backflow is higher.  \end{abstract}
 
\date{\today}
\maketitle

\section{INTRODUCTION}

Manipulation of heat at the microscopic level has been intensively studied in recent years. We can find several proposals of heat engines and/or heat pumps which consist of a finite dimensional quantum system coupled to multiple environments \cite{Wang,Hanggi,Li,Kosloff,Segal}. In these works, environmental effects have mainly been described using the formalism of the master equation in the well-known Gorini-Kossakowski-Sudarshan-Lindblad (GKSL) form \cite{GKSL}, suitably \Brev{embedded} in the context of full counting statistics (FCS) \cite{Esposito}, where the dynamics is described by a semigroup therefore enabling only simplified exchanges of energy. 
Such master equations have been proven to describe quite \Brev{faithfully} many physical open systems in the Born-Markov approximation, which, however, represents a restrictive requirement that often fails to provide an accurate description of the dynamics at short and intermediate time scales, especially in the presence of a structured environment \cite{Liu2011NAT,Huelga2013}.
Therefore, the importance of considering generalized master equations with time-dependent coefficients, also spurred by the recent developments of quantum technologies in short-time and/or low-temperature regions \cite{Nibbering,Saikan,Woggon,Madsen}, has became a leitmotif in this field \cite{Esposito, Flindt, Braggio}. 
Within this more realistic framework, however, a non-Markovian description of the dynamics is necessary in order to correctly characterize the system. Due to this fact, much effort has been devoted in the last decade to formally define and quantify the degree of non-Markovianity of a given dynamics \cite{Wolf2008PRL,Breuer2009PRL,Breuer2012JPB,Rivas2010PRL,Lu2010PRA,
LuoPRA2012,Lorenzo2013PRA,Bylicka2013arxiv,Chruscinski2014PRL,Rivas2014,
Breuer2015}, as well as developing reservoir-engineering techniques to manipulate it \cite{Verstraete,Biercruk,Liu2011NAT}.
\GGrev{Note that in the present paper, following recent work, see e.g. \cite{Rivas2014,Breuer2015} for an overview, we allow the term non-Markovian also for dynamics described by time-local equations, provided the coefficients in the equations become at least temporarily negative, thus allowing for revivals in the behavior of certain system observables. 
Such features that are considered in the paper are related to a competition between system and environment relaxation time-scales.}
One of the most active research areas in this field is nowadays dedicated to the challenge of exploiting non-Markovianity as a resource to improve, for example, quantum communication protocols \cite{Laine2014,Paris}. 

Our work is inserted in this framework and in particular focuses on the study \BVrev{of the behavior in time} of the energy flow between a system of interest and its environment, quantified through FCS methods, without using the full Born-Markov approximation and only assuming a second-order time-convolutionless expansion of the generator of the dynamics, that is working in the Born approximation. In this scenario, at variance with what happens in the \BVrev{Born-Markov}(semigroup) regime, where a one-way-only energy current can occur, non-Markovian dynamical effects generally arise and the rate by which system and environment exchange energy can oscillate in time and energy can even come back from the environment to the system. 
In order to capture this phenomenon, we introduce a condition and a suitable quantifier of the energy backflow.

We then illustrate this scenario considering a spin-boson model, which provides a paradigm in the description of dissipative two-level systems and has found wide applicability in many important situations (see, e.g., \cite{Weiss,Leggett1987}).
Finally, we study the connection between the introduced quantifier for the energy backflow and the occurrence of non-Markovianity as introduced by Breuer, Laine, and Piilo in \cite{Breuer2009PRL}.

The present paper is organized as follows. In Sec. \ref{sec:Formalism} we recall the FCS formalism and we apply it to the study of energy transfer between an open quantum system and its environment. We also give, in Sec. \ref{sec:FormalismB}, a condition and a quantitative measure for the occurrence of energy backflow. We apply this construction to the spin-boson model in Sec. \ref{sec:SB}, where we illustrate the system and discuss the results in detail. Finally in Sec. \ref{sec:Markov} we build a connection between the occurrence of energy backflow and of non-Markovianity. Conclusions are drawn in Sec. \ref{sec:Conclusions}.

\section{FORMALISM}
\label{sec:Formalism}

In this section, we make an overview of the principal aspects of the \GGrev{FCS} formalism which will be necessary to our purposes. In Sec. \ref{sec:FormalismB} we proceed to introduce the concept of energy backflow within \BVrev{this} framework.

\subsection{Full counting statistics}

Let us start by briefly recalling the FCS formalism as developed in Esposito, Harbola, and Mukamel \cite{Esposito}.  Consider a system of interest interacting with its environment according to the Hamiltonian $\mathcal{H}=\mathcal{H}_0+\mathcal{H}_{\rm int}$, where $\mathcal{H}_0=\mathcal{H}_S+\mathcal{H}_E$ denotes the sum of the system Hamiltonian $\mathcal{H}_S$ and of the environmental Hamiltonian $\mathcal{H}_E$, while $\mathcal{H}_{\rm int}$ is the interaction Hamiltonian.  Within this framework, the time evolution of an environmental observable $Q$ which is transferred from the relevant system into the environment (or vice versa) is reconstructed by looking at the difference in the outcomes of two-point projective measurement.  Let $q_0$ and $q_t$ denote such outcomes, respectively, at the initial time $t=0$ and at a later time $t$, of the measurement of the environmental observable $Q$. The joint probability to have \GGrev{obtained} these outcomes reads
\begin{equation} 
P\left[q_t,q_0\right] \!=\! \mathrm{Tr}_{S+E}\left\{\Pi_{q_t}U(t,0)\Pi_{q_0}\rho_{SE}(0) \Pi_{q_0} U^{\dagger}(t,0)\Pi_{q_t}\right\}, 
\end{equation} 
where $\Pi_{q_t} = \ket{q_t}\bra{q_t}$ indicates the projective measurements relative to eigenvalue $q_t$, while $U(t,0)$ denotes the total unitary evolution operator and $\rho_{SE}(0)$ is the initial state of the composite system. The information on the variation of the observable $Q$ is then encapsulated in the probability density 
\begin{equation} 
P_t(\Delta q) = \sum_{q_t,q_0} \delta\left(\Delta q - (q_t-q_0)\right) P\left[q_t,q_0\right], 
\end{equation} 
where $\delta(\cdot)$ stands for the Dirac $\delta$ distribution.  Introducing the cumulant-generating function 
\begin{equation} 
S_t(\chi) = \ln \int_{-\infty}^{+\infty}\,d(\Delta q) P_t(\Delta q) e^{i\chi\Delta q},
 \end{equation} 
where $\chi$ is often referred to as the counting field, the \BVrev{$n$th} cumulant of the probability distribution $P_t(\Delta q)$ is then readily given by the \BVrev{$n$th} derivative of $ S_t(\chi) $:
\begin{equation} 
\label{cumulantFCS}
\mean{\Delta q^n}_t = \left.\frac{\partial^n S_t(\chi)}{\partial (i\chi)^n}\right|_{\chi=0}.  
\end{equation}
This procedure can be directly applied in the context of open quantum systems as follows. If we assume that $\left[\Pi_{q_0},\rho_{SE}(0)\right]=0$ and moreover use the fact that, if the spectral decomposition of $Q$ is $Q = \sum_{q} q \Pi_q$, then $f(Q) = \sum_{q} f(q) \Pi_q$ holds for any function $f$, we readily obtain the following relation:
\begin{equation}
\sum_{q_0} e^{-i\chi q_0} \Pi_{q_0} \rho_{SE}(0) \Pi_{q_0} = e^{-i\frac{\chi}{2} Q(0)} \rho_{SE}(0) e^{-i\frac{\chi}{2} Q(0)}.
\end{equation}
By means of it, it is straightforward to prove that the cumulant-generating function can be written as
\begin{equation}
S_t(\chi) = \ln\mathrm{Tr}_{S}\left\{\rho^{\chi}(t)\right\},
\end{equation}
where we have introduced the conditional density operator
\begin{equation}
  \label{eq:3}
 \rho^{\chi}(t) \equiv \mathrm{Tr}_{E} \left\{U_{\chi/2}(t,0) \rho_{SE}(0) U^{\dagger}_{-\chi/2}(t,0)\right\},
\end{equation}
which evolves according to the modified evolution operator 
$
U_{\chi}(t,0) \equiv e^{i\chi Q(t)} U(t,0) e^{-i\chi Q(0)}.
$
Obviously, for $\chi=0$ we retrieve the usual evolution operator $U_{\chi=0}(t,0) = U(t,0) $ and the statistical operator $\rho^{\chi=0}(t) = \rho(t)$.

\subsection{Energy backflow}
\label{sec:FormalismB}

Let us apply the FCS formalism to the study of energy transfer between a system and its environment. 
To this aim we select the observable $Q$ as the environmental energy, i.e., $ Q \equiv \Ham_E $, whose spectrum we assume to be time-independent since no external driving fields are considered. 
In the absence of initial correlations between the system of interest and its environment, i.e. ,
\begin{equation}\label{initialfactstate}
\rho_{SE}(0) = \rho_S(0)\otimes\rho_E,
\end{equation}
with $\rho_E$ being a Gibbs state relative to the temperature $T_E$, the evolution of the modified operator $\rho^{\chi}(t)$ is given in terms of the time-convolutionless generalized master equation (GME) \cite{Kubo,TCL2,Shibata}
\begin{equation}\label{MEform}
\frac{d}{dt}\rho^{\chi}(t) = \Xi^{\chi}(t)\rho^{\chi}(t),
\end{equation}
where the time-dependent superoperator $\Xi^{\chi}(t) $ in the second-order approximation has the form
\begin{align}\label{eq:2TCLchi}
&\Xi^{\chi}(t)\left[\omega\right] =\BVrev{-i}\left[\Ham_S,\omega\right]\notag\\
&- \int_0^t \, d\tau \,\mathrm{Tr}_{E} 
\left[{\Ham_{int}},\,\left[{\Ham_{int}}(-\tau),\,\omega\otimes\rho_E(0)\right]_{\chi}\right]_{\chi}
\end{align}
where $\left[\CUrev{\Ham_{int}(t)} \,,\, B\right]_{\chi} \equiv \CUrev{\Ham_{int}^{\chi}(t)}B - B \CUrev{\Ham_{int}^{-\chi}(t)}$, with $\CUrev{\Ham_{int}^{\chi}(t) = e^{(i/2) \chi \GGrev{\Ham_E}} \Ham_{int}(t)  e^{-(i/2) \chi \GGrev{\Ham_E}}}$ and $\CUrev{\Ham_{int}}(t) = e^{i\Ham_0 t} \CUrev{\Ham_{int}} e^{-i\Ham_0 t}$.
In the expressions above and in the remainder of the work we set $\hbar=1$ for simplicity.
The formal solution of \eqref{eq:2TCLchi} has the form
\begin{equation}
\ket{\rho^{\chi}(t)} = T_+ \exp\left[\int_0^t\,d\tau \bgreek{\Xi}^{\chi}(\tau)\right]\ket{\rho(0)},
\end{equation}
with  $T_+$ indicating the chronological time ordering operator and $\ket{\rho^{\chi}(t)}$ and $\bgreek{\Xi}^{\chi}(t)$ denoting, respectively, the vector and matrix forms in the Hilbert-Schmidt space of the operator $\rho^{\chi}(t)$ and of the superoperator $\Xi^{\chi}(t)$.

The time-dependent first moment of the energy transfer can be consequently expressed, using \eqref{cumulantFCS}, as \cite{UchiyamaPRE}
\BVrev{
\begin{equation}\label{firstmoment1}
\langle \Delta q \rangle_t = \bra{1} \frac{\partial}{\partial (i\chi)}\ket{\rho^{\chi}(t)}_{|\chi=0},
\end{equation} 
}
where $\bra{1}$ denotes the trace operation in Hilbert-Schmidt space. 
Since the density operator satisfies the relation $\mathrm{Tr}_{S} \{\rho^{\chi=0}(t)\}=1$, the state $\bra{1}$ is a left
eigenstate of $\bgreek{\Xi}^{\chi}(t)$ relative to the zero eigenvalue. Using this relation, Eq. \eqref{firstmoment1} can be expressed in the more compact form
\begin{equation}\label{EnFluxGeneral}
\langle \Delta q \rangle_t = \int_0^t\,d\tau \theta(\tau),
\end{equation}
where we have introduced the function
\BVrev{
\begin{equation}\label{thetaGen}
\theta(t) \equiv\bra{1}\frac{\partial \bgreek{\Xi}^{\chi}(t)}{\partial (i\chi)}\ket{\rho (t)}_{|\chi=0},
\end{equation}
}
which describes the energy flow per unit of time, i.e., the rate of energy \GGrev{exchanged} between the system of interest and its environment.

In the remainder of the paper $\theta(t)$ will be the crucial quantity under investigation. 
The rate by which the system and its environment exchange energy may, \Brev{in fact, strongly vary} depending on the several parameters characterizing the dynamics. In particular, adopting the terminology from the context of non-Markovian dynamics \cite{Breuer2009PRL,Fanchini2014PRL}, we speak of regions of \textit{energy backflow} from the environment to the system whenever, considering situations
which in the Born-Markov semigroup approximation would lead to a non-negative steady energy transfer from system to environment, we have that at some time $t$
\BVrev{
\begin{equation}\label{condenb}
\theta(t) < 0.
\end{equation}
}
Building on this condition, a measure for the total amount of energy which has flown back from the environment to the system during the evolution is naturally introduced as
\begin{equation}\label{enbackflow}
\mean{\Delta q}_{back} = \max_{\rho_S(0)}\,\frac{1}{2} \int_0^{+\infty}\,dt\, \left(\left|\theta(t)\right|-\theta(t)\right),
\end{equation}
where the maximization procedure is performed over all possible initial states of the reduced system.
Note that the integrand of \eqref{enbackflow} is different from zero if and only if $\theta(t)$ assumes negative values. 
Moreover, it represents, in principle, a measurable quantity. 
However, despite the formal similarity between this quantifier and some of the recently introduced non-Markovianity measures \cite{Breuer2009PRL,Rivas2010PRL,Fanchini2014PRL}, it should not be confused with an alternative non-Markovianity measure, rather providing only an estimate of the energy backflow.

\section{The Spin-Boson Model}
\label{sec:SB}

In the present section we study the energy transfer in the so-called spin-boson model. 
This model, which describes a spin-$\frac{1}{2}$ system interacting with an environment consisting of an infinite number of bosonic modes, has been studied extensively over the past decades since it represents the prototypical model of dissipative two-level system \cite{Weiss,Leggett1987}.
Let us start from the Hamiltonian of the composite system, which reads \CUrev{$\mathcal{H}=\mathcal{H}_S +\mathcal{H}_E+\mathcal{H}_{\rm int}$,} with
\begin{equation}\label{eq:Hamiltonian}
\CUrev{\mathcal{H}_{S}=\frac{\omega_0}{2}\sigma_z, \quad \mathcal{H}_{E}=\sum_k\omega_kb^{\dagger}_k b_k, \GGrev{\quad\text{and}\quad}\mathcal{H}_{\rm int}= \sigma_x \otimes B_E},
\end{equation}
where $\sigma_{z,x}$ denote the usual Pauli matrices, $\omega_0$ is the energy difference between excited ($|1\rangle$) and ground ($|0\rangle$) state of the system, \GGrev{$\omega_k$ stands for the energy of the $k$th bosonic mode, and $g_k$ is the coupling strength between the mode and the system}. Finally, in \eqref{eq:Hamiltonian}, we have denoted
\begin{equation}\label{Be}
B_E \equiv \sum_k \left( g_k b^{\dagger}_k + g^*_k b_k \right),
\end{equation}
with $ b_k $ and $ b_k^\dagger$  being the bosonic annihilation and creation operator of the environment relative to mode $k$. 

A GME of the form \eqref{MEform} can be obtained for this model at second order in a perturbation expansion \cite{UchiyamaPRE}, whose analytical solution can be approached moving to the Hilbert-Schmidt space. In this space, the conditional density operator $\rho^{\chi}(t)$ is represented by the vector $\ket{\rho^{\chi}(t)} = \left(\rho_{00}^{\chi}(t),\rho_{01}^{\chi}(t),\rho_{10}^{\chi}(t),\rho_{11}^{\chi}(t)\right)^T$, where 
\begin{equation}
\rho_{\alpha}^{\chi}(t) = \mathrm{Tr}_{S} \left\{\sigma_\alpha^{\dagger}\rho^{\chi}(t)\right\}
\end{equation}
and $\lbrace\sigma_\alpha\rbrace_{\alpha=0,1,2,3} = \lbrace \ket{0}\bra{0}, \ket{0}\bra{1},\ket{1}\bra{0},\ket{1}\bra{1}\rbrace$.
Correspondingly, the super-operator $\Xi^{\chi}(t)$, now regarded as a linear map on the space of linear operators on $ \mathbb{C}^2 $, is \Brev{given} by a $4\times4$ matrix $\bgreek{\Xi}^{\chi}(t)$, whose entries are explicitly given by \cite{UchiyamaPRE}
\begin{equation}\label{matrixXichi}
\bgreek{\Xi}^{\chi}(t) = -\int_0^t\,d\tau\, \begin{pmatrix}
V_+(\tau) & 0 & 0 & W_+^{\chi}(\tau) \\ 
0 & Y_+(\tau) & Z_+^{\chi} & 0 \\
0 & Z_-^{\chi}(\tau) & Y_-(\tau) & 0 \\
W_-^{\chi}(\tau) & 0 & 0 & V_-(\tau)
\end{pmatrix}.
\end{equation}
\GGrev{The quantities appearing in \eqref{matrixXichi} are linear combinations of the environmental correlation function
\begin{equation}\label{envcorrfunc}
\Phi(\tau) \equiv\mathrm{Tr}_{E} \left\{B_E B_E(-\tau)\rho_E\right\},
\end{equation}
defined by}
\begin{align}\label{eqB6B9}
V_{\pm}(\tau) &= \Phi(\tau)e^{\mp i\omega_0\tau}+\Phi(-\tau)e^{\pm i\omega_0\tau}, \notag\\
W^{\chi}_{\pm}(\tau) &= -\left[\Phi(\tau-\chi)e^{\pm i\omega_0\tau}+\Phi(-\tau-\chi)e^{\mp i\omega_0\tau}\right],\notag\\
Y_{\pm}(\tau) &= 2 Re\left[\Phi(\tau)\right] e^{\mp i \omega_0 \tau},\notag\\
Z^{\chi}_{\pm}(\tau) &=  -\left[\Phi(\tau-\chi)+\Phi(-\tau-\chi)\right]e^{\pm i\omega_0\tau}.
\end{align}
We stress here that the familiar master equation describing the evolution of the statistical operator in the spin-boson model \cite{ClosBreuer} can be obtained from \eqref{matrixXichi} simply by setting the counting field parameter $\chi=0$. 

A crucial role in the definition of $\Phi(t)$ is played by the spectral density
$J(\omega) = \sum_k |g_k|^2 \delta(\omega-\omega_k)$,
which describes both the distribution of bath modes and their interaction strength with the system.
In the limit of a continuous distribution of environmental modes the spectral density can be described by a smooth function which we take of the form
\begin{equation}\label{SpectralDensity}
J(\omega) = \lambda \omega e^{-\frac{\omega}{\Omega}},
\end{equation}
which shows an Ohmic behavior at low frequencies, a linear dependence on the coupling strength $\lambda$, and finally an exponential cutoff part.
The analytic form for the environmental correlation function in this case can be found and reads
\begin{align}\label{envcorrcont}
\Phi(\tau) &= \int_0^{+\infty} d\omega\, J(\omega) \left[\coth\left(\frac{\omega}{2T_E}\right)\cos(\omega\tau)-i\sin(\omega\tau)\right] \notag\\
&\equiv \frac{1}{2} \left(D_1(\tau) - i D_2(\tau)\right)
\end{align}
where $T_E$ denotes the environmental temperature, the Boltzmann and Planck constants have been set equal to one $k_B = \hbar = 1$, and the functions $D_1(\tau)$ and $D_2(\tau)$, respectively known as \textit{noise} and \textit{dissipation} kernels \cite{Breuer2002}, have the concrete expressions
\begin{align}\label{eq:D1D2}
&D_1(\tau) = 2\int_0^{+\infty} d\omega J_{eff}(\omega,\Omega,T_E) \cos(\omega\tau) \notag\\
&\!=\! 2\lambda\left[\Omega^2\frac{(\Omega\tau)^2-1}{(1+(\Omega\tau)^2)^2}\!+\! 2 T_E^2\, Re\left[\psi' \left(\frac{T_E(1+i\Omega\tau)}{\Omega}\right)\right]\right] 
\nonumber
\\
&D_2(\tau) = 2\int_0^{+\infty} d\omega J(\omega) \sin(\omega\tau)=\frac{4\lambda\Omega^3 \tau}{(1+(\Omega\tau)^2)^2},
\end{align}
with ${\psi'(z)} $ being the derivative of the Euler digamma function $\psi(z) = {\Gamma'(z)}/{\Gamma(z)}$ and where we have introduced the
effective spectral density in the noise kernel
\begin{equation}\label{Jeff}
J_{eff}(\omega,\Omega,T_E) \equiv J(\omega) \coth\left(\frac{\omega}{2T_E}\right).
\end{equation}
Substitution of expression \eqref{matrixXichi} of the superoperator $\bgreek{\Xi}^{\chi}(t)$ in Eq. \eqref{thetaGen} leads to the following expression \GGrev{for} the energy flow per unit of time
\begin{equation}\label{theta}
\theta(t) = \left[ w_+(t)-w_-(t)\right]\rho_{00}(t)-w_+(t),
\end{equation}
where we have defined the quantity
\begin{equation}\label{wpm}
w_{\pm}(\tau) \equiv \left.\frac{\partial }{\partial (i\chi)} \int_0^{\tau}\,ds\, W_{\pm}^{\chi}(s)\right|_{\chi=0}.
\end{equation}
It is clear from \eqref{theta} that only the populations of the open system contribute to the energy flow. 
However, it is interesting to look in more detail at the contributions appearing in Eq. \eqref{theta}.
Building on the results detailed in Appendix \ref{app:AppB}, the right-hand side of Eq. \eqref{theta} can be reexpressed in the form
\begin{equation}\label{relationimp}
\theta(t) = \omega_0 \frac{d}{dt}\rho_{00}(t) + f(t),
\end{equation} 
where
\begin{equation}
f(t) \equiv -\delta p(t) D_1(t)\sin(\omega_0 t) + D_2(t)\cos(\omega_0 t),
\end{equation}
having introduced the difference in the system's populations $\delta p(t)\equiv \rho_{11}(t)-\rho_{00}(t) $.
The first term on the right-hand side of Eq. \eqref{relationimp} in fact just corresponds to the time derivative of the change in the \GGrev{free} system's energy, since it is proportional through the fundamental system energy $\omega_0$ (we remind that $\hbar=1$) to the fraction of the system's population that moves to the ground state. 
The second term, $f(t)$, is instead a combination of elementary oscillating functions and environmental kernels: The \BVrev{first contribution} is driven by the noise kernel $D_1(t)$ and also depends on the solution for the ground-state population of the system $\rho_{00}(t)$, at variance with the \BVrev{second} which is proportional only to the dissipation kernel, therefore also being independent of the temperature of the bath.

The integral form of \eqref{relationimp} can also be considered
\begin{equation}\label{relationimpint}
\langle\Delta q\rangle_t = \omega_0\left(\rho_{00}(t)-\rho_{00}(0)\right) + F(t),
\end{equation}
where $F(t) = \int_0^t\,d\tau f(\tau)$.
Equation \eqref{relationimpint} shows that the variation in the environmental energy, obtained in this case as the FCS mean $\mean{\Delta q}_t$, is given by the sum of two distinct contributions. The first term on the right-hand side corresponds to the variation of the reduced system's energy, but there is an additional contribution which depends both on the detailed structure of the environment, and on the coupling between system and environment through the dissipation and noise \GGrev{kernels}. In the long-time limit, which corresponds to the Born-Markov approximation, this additional contribution vanishes since $f(t)$ goes to zero according to the behavior of both dissipation and noise kernel as given by \eqref{eq:D1D2}; see Appendix \ref{app:AppA} for details. 

\subsection{Numerical results and discussion}

In this section we illustrate and discuss the results of the numerical evaluation of the measure of energy \GGrev{backflow \eqref{enbackflow} for the considered model} for different values of the relevant parameters. 

\begin{figure}[!ht]
{\bf (a)}
\\
\includegraphics[width=0.84\columnwidth]{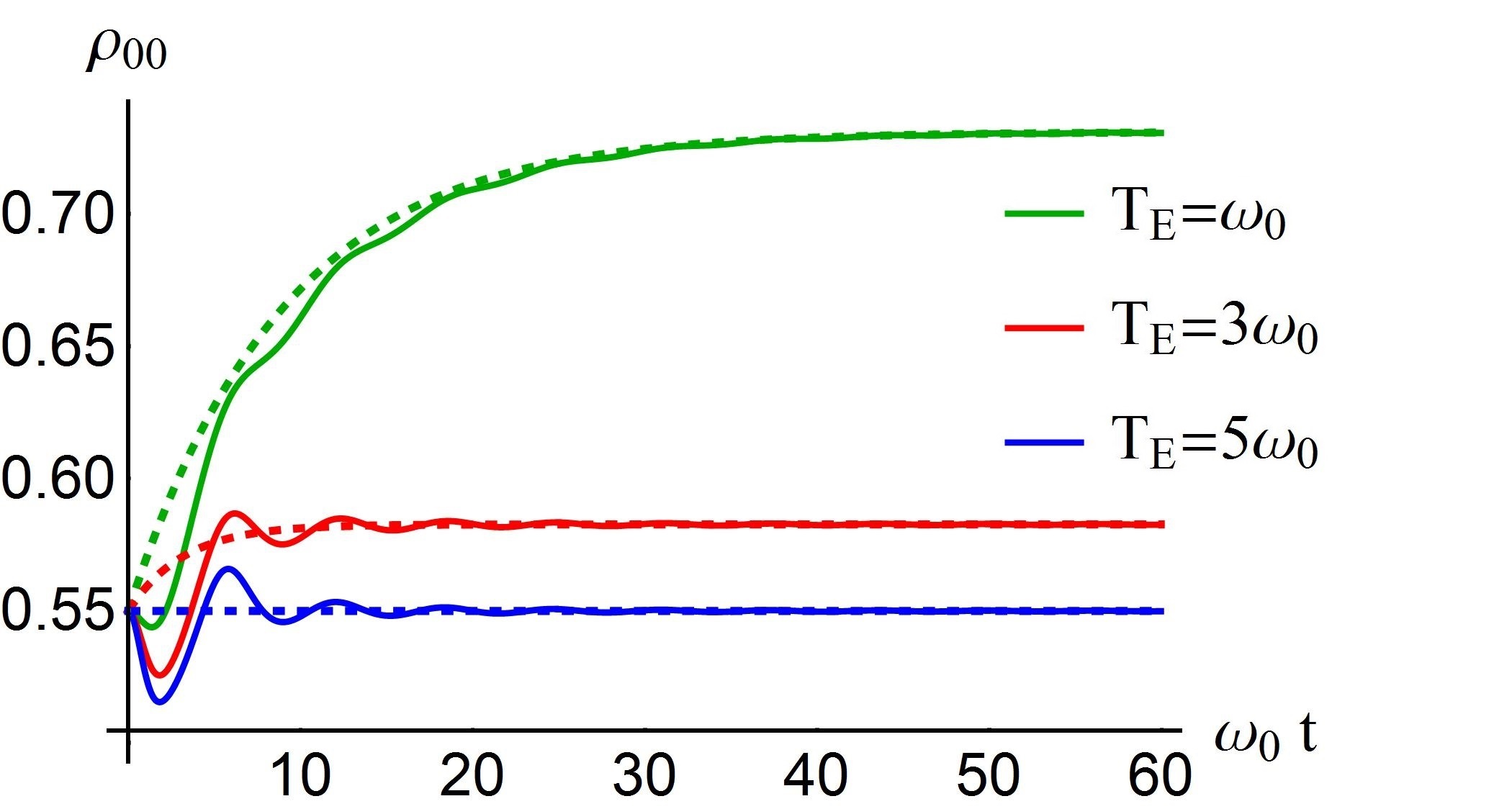}
\vspace{.5truecm}
\\
{\bf (b)}
\\
\hspace*{-.98truecm}
\includegraphics[width=0.84\columnwidth]{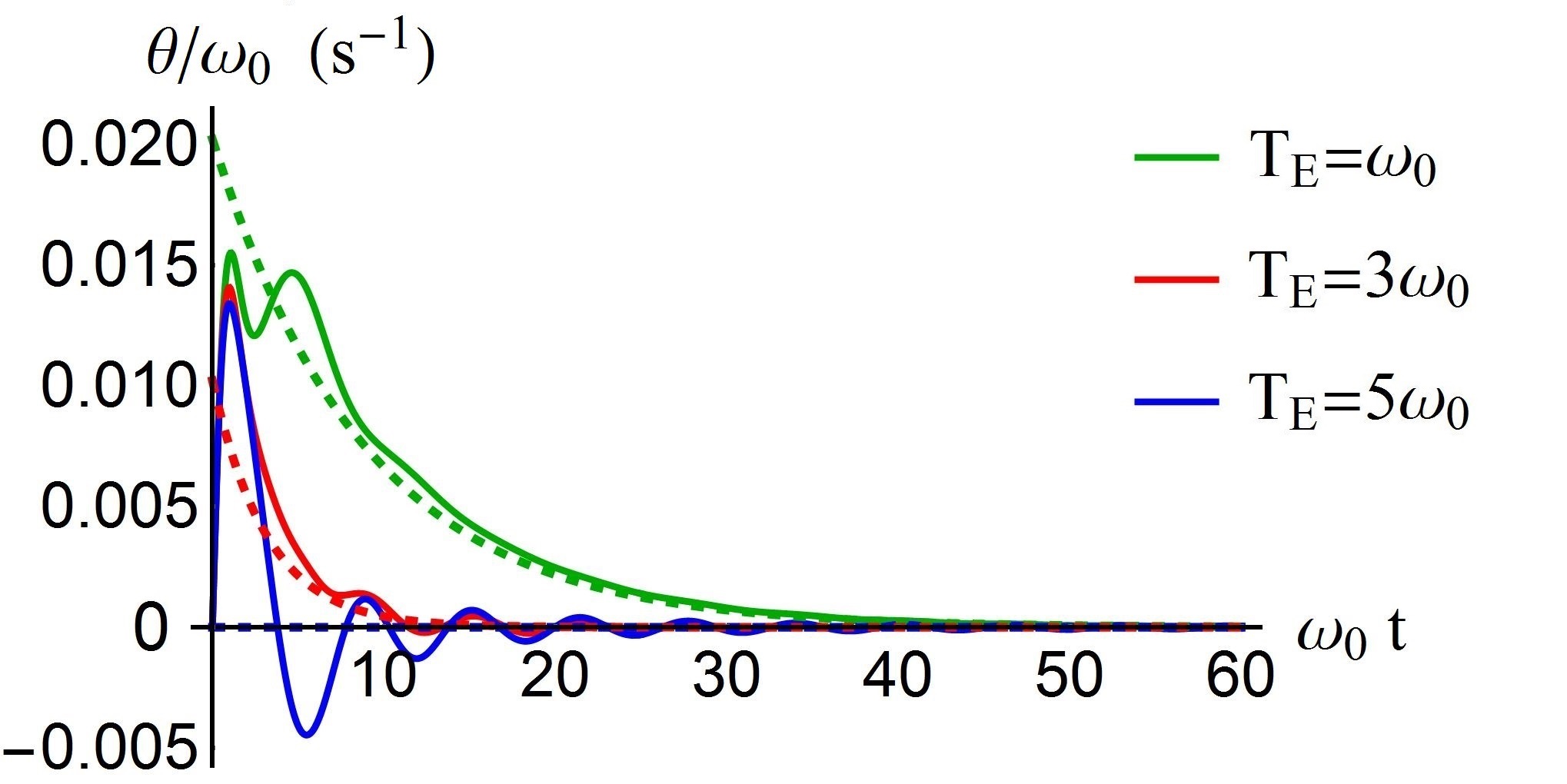}
\caption{(Color online) {\bf (a)}  Time evolution of the ground-state population $\rho_{00}(t)$ for $\Omega=0.4\omega_0 $, $\lambda=0.1$, and $T_S=5\omega_0$, for different values of the environmental temperature $T_E/\omega_0=1,\,3,\,5$. The dashed lines refer to the Born-Markov approximation, while the solid lines refer to the time evolution obtained by the time-convolutioness GME. {\bf (b)} Time evolution of $\theta(t) / \omega_0$ ($\mathrm{s}^{-1}$) for the same parameters values and specific choice of initial Gibbs states. One can notice that above a certain temperature gradient between system and environment the energy backflow disappears.}
\label{fig1}
\end{figure}

\begin{figure}[!ht]
\includegraphics[width=\columnwidth]{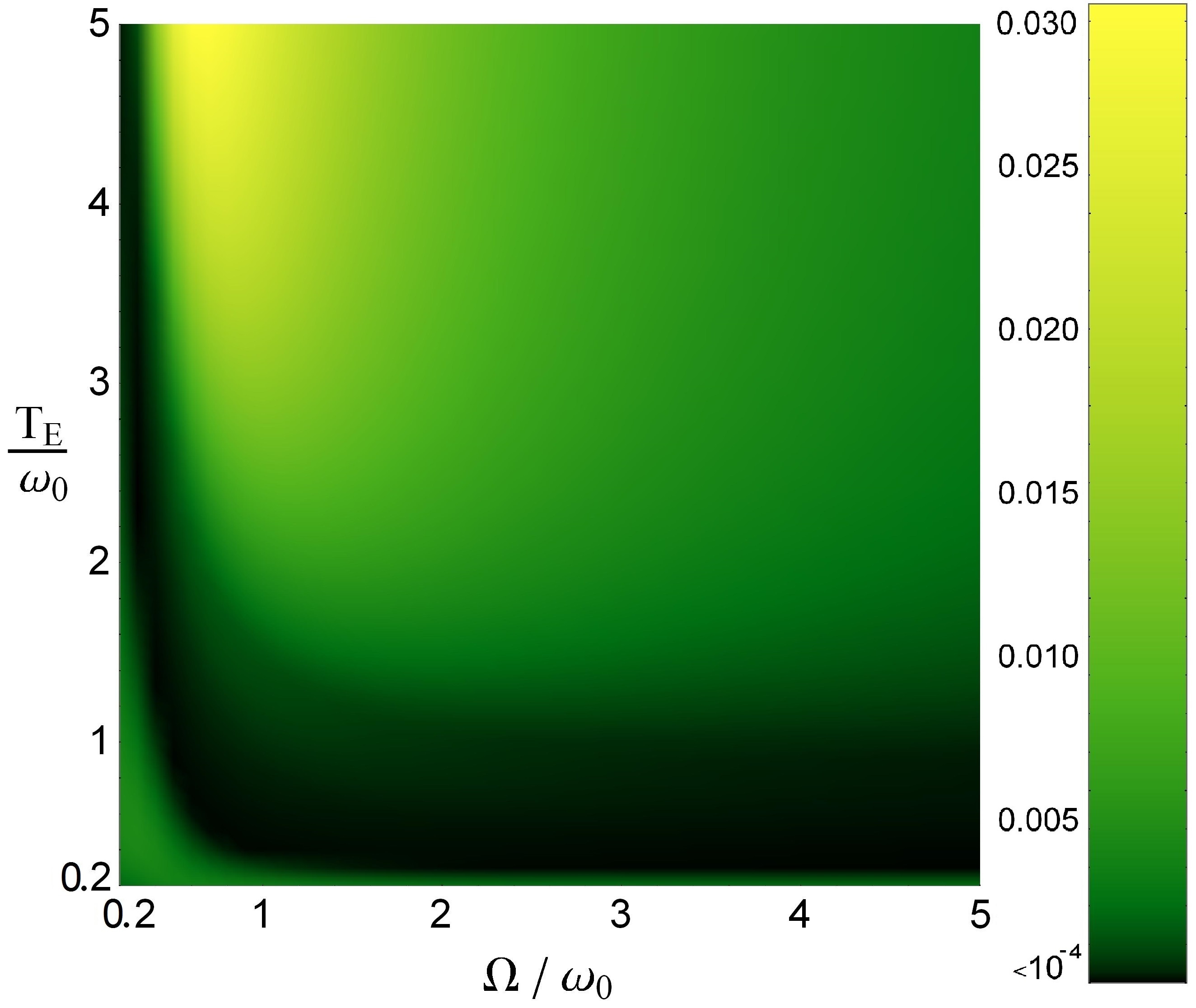}
\caption{(Color online)  Plot of the energy backflow $\mean{\Delta q}_{back}(\Omega,T_E)$ in units of $\omega_0$ as given by Eq. \eqref{enbackflow}, for values of the parameters $\left( \Omega/\omega_0, T_E/\omega_0\right)$ in the range $(0.2 , 5) \times (0.2 , 5)$. The coupling constant is chosen to be $\lambda=0.1$ in conformity with the choice for the non-Markovianity measure in \cite{ClosBreuer}; see also Fig. \ref{fig:nM} in Appendix \ref{app:NM}. The upper limit of time integration has been chosen to be equal to $100 \omega_0^{-1}$. The effective system temperature $T_S$ has been chosen equal to $T_E$ in order to maximize the oscillations of $\theta(t)$.}
\label{fig:ebf}
\end{figure}

In the remainder of the work we consider the following form for the initial state of the reduced system
\begin{equation}\label{initialchoice}
\rho_S(0) = Z^{-1}\left(\ket{0}\bra{0}+e^{-\omega_0/T_S}\ket{1}\bra{1}\right) \,,\quad Z=1+e^{-\omega_0/T_S},
\end{equation}
which is a Gibbs state with an effective temperature $T_S$, here chosen to be greater than or equal to the environmental one.  Indeed, in order to study the flow of energy from the environment back
to the system, we are interested in situations
which in the Born-Markov approximation, corresponding to a semigroup description, would lead to a steady energy transfer from system to environment. This situation corresponds to setting $T_S \ge T_E$, so that in this case one can properly speak of energy backflow. Figure \ref{fig1} shows the time evolution of the ground-state population $\rho_{00}(t)$ \textbf{(a)} and the energy flow per unit of time $\theta(t)$ \Brev{as given by \eqref{theta}} \textbf{(b)} in the weak-coupling limit $\lambda=0.1$ and in units of $\omega_0$, for $\Omega=0.4\omega_0$, $T_S=5\omega_0$, and different values of the environmental temperature $T_E/\omega_0=1,\,3,\,5$.
We refer to Appendix \ref{app:AppA} for details about the differential equation obeyed by $\rho_{00}(t)$ as well as its formal solution, which is plotted here.
Solid lines in Fig. \ref{fig1} refer to the solutions obtained from the second-order \Brev{time-convolutionless} expansion of the generator, while the dashed lines denote the ones obtained in the Born-Markov approximation. 
It is clear from these plots that the time behavior of the solution of the ground-state population, $\rho_{00}(t)$, is related to the time behavior of the energy flow per unit of time $\theta(t)$: Both quantities, in fact, show a transition from oscillating to monotone behavior at almost the same time.
We find that the oscillations of the exact solution (solid lines) of both quantities almost disappear in the long-time limit and superimpose the asymptotic value determined by the Born-Markov approximated solutions (dashed lines). The markedly different behaviors of solid and dashed lines in short and intermediate time, however, neatly show the inadequacy of Born-Markov approximation apart from the long-time limit case. 
\GGrev{An interesting property of the energy flow is represented by the first positive peak of $\theta(t)$, which can be observed even when the initial temperatures of the reduced system and of the environment are equal to each other; see Fig. \ref{fig1}\textbf{(b)}. Such peak is a general feature due to choice of the initial factorized state \eqref{initialfactstate}, which is essential in order to have a well-defined dynamical map \cite{Breuer2002}, but represents a non-equilibrium preparation
\begin{equation}
\rho_{SE}(0) = \frac{e^{-\Ham_S/T}}{Z_S} \otimes \frac{e^{-\Ham_E/T}}{Z_E} \neq \frac{e^{-\Ham/T}}{Z},
\end{equation}
with $Z_S$ and $Z_E$ being the partition functions of the reduced system and environment respectively and $Z$ being the partition function of the composite system $S+E$.  This factorized non-equilibrium initial preparation is known to lead \cite{AnkerholdPRB2014,SchmidtPRB2015} to an energy exchange between system and environment which takes place on short time scales due to the establishment of proper system-environment equilibrium correlations.} Moreover, it can be noticed from Fig. \ref{fig1} \textbf{(b)} how the value of the first local minimum of $\theta(t)$ decreases for decreasing values of the difference $T_E - T_S$, attaining its lowest value for $T_E = T_S$.  Strong numerical evidences suggest that this trend is maintained for
all values of the relevant parameters $\lambda,\Omega,T_E$, thus making it possible to conclude that energy backflow \eqref{enbackflow} [i.e., the area of the negative region of $\theta(t)$] is maximized by the choice of having initial system and environment at the same temperature, \Brev{which can be understood considering the fact that in this case there is no initial temperature gradient.}
We note that the choice \eqref{initialchoice} for the class of initial system's states does not affect the validity of this result. In fact, since the equations of motion for the coherences and populations are decoupled from each other and since the coherences do not enter the expression of $\theta(t)$, any initial state with nonzero coherence is equivalent for this purpose to a diagonal state, which can always be recast in a Gibbs form \eqref{initialchoice} relative to an effective temperature $T_S$.
We have thus evaluated the amount of energy backflow, as estimated by Eq. \eqref{enbackflow}; the result $\mean{\Delta q}_{back}(\Omega,T_E)$ is given in Fig. \ref{fig:ebf}, for the value of the coupling strength $\lambda=0.1$ and for values of the parameters $\left( \Omega/\omega_0, T_E/\omega_0\right)$ in the range $(0.2 , 5) \times (0.2 , 5)$.
We remark that the values of the amount of energy backflow, given in units of $\omega_0$, are represented on a color-bar scale for better visualization.

The calculation has been explicitly carried out by numerically evaluating the integral \eqref{enbackflow} over a fine grid of 2500 points. The maximization over the initial system state 
has been performed by setting the effective temperature $T_S$ of the system to be equal to the environmental one $T_E$.
Moreover, the upper limit in the integral \eqref{enbackflow} has been chosen to be equal to $100 \omega_0^{-1}$: After such time interval, in fact, the energy flow per unit of time $\theta(t)$ superimposes, for this value of the coupling strength, the Born-Markov solution, i.e., oscillation of $\theta(t)$ as well as negativity regions are no longer significant.

\begin{figure}[!ht]
{\bf (a)}
\\
\includegraphics[width=.73\columnwidth]{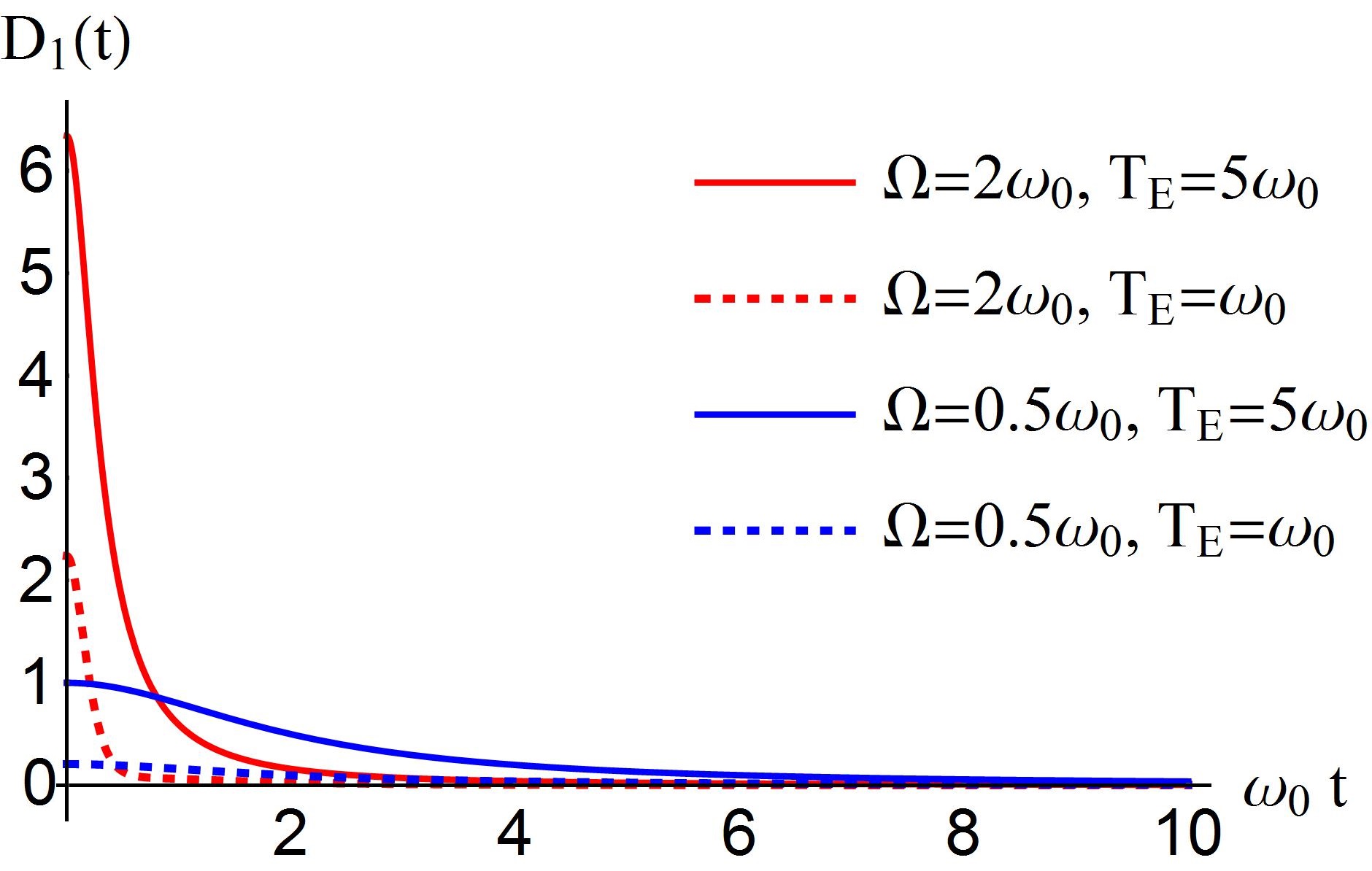}
\vspace{.1truecm}
\\
{\bf (b)}
\\
\includegraphics[width=.73\columnwidth]{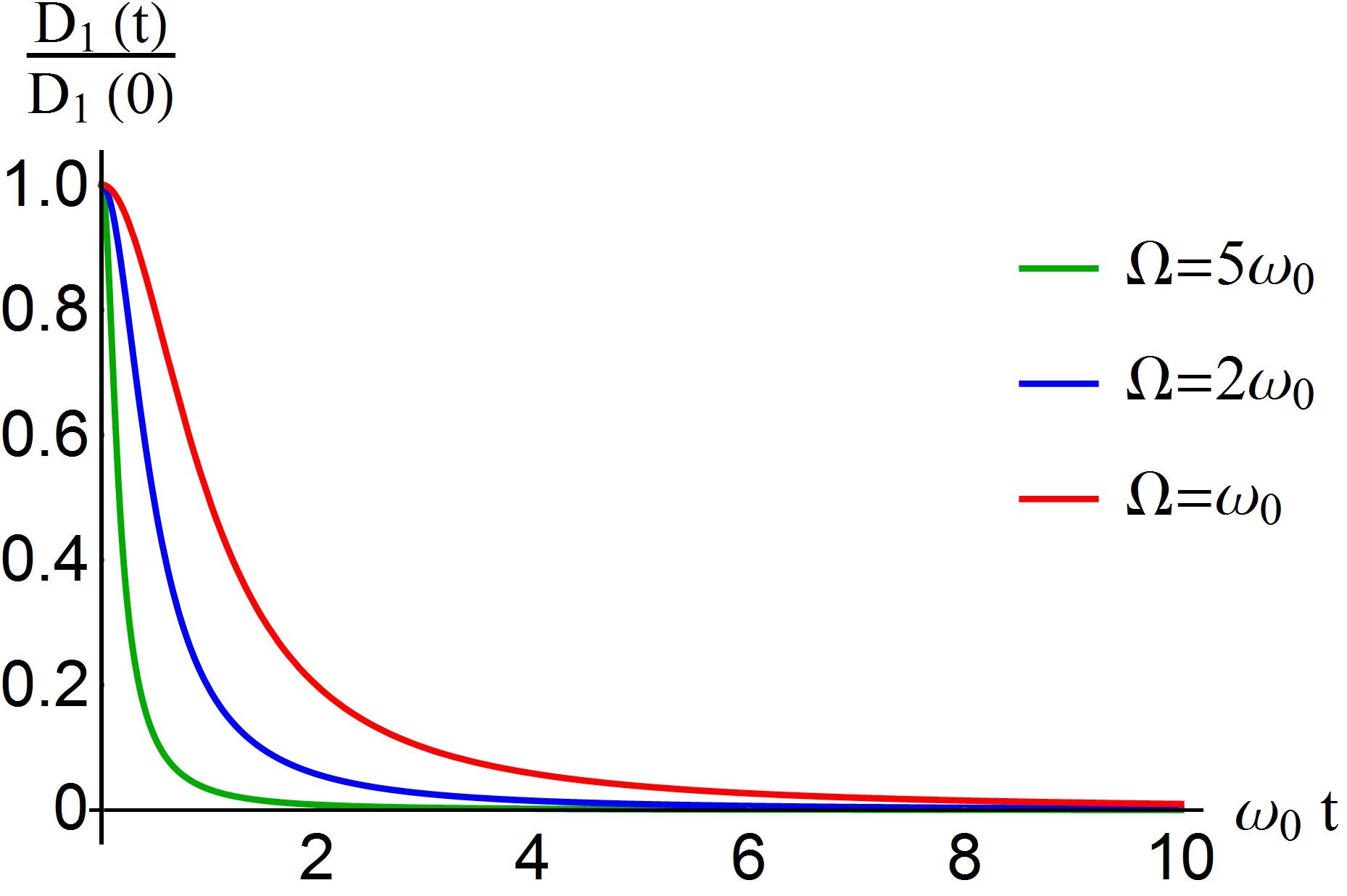}
\vspace{.1truecm}
\\
{\bf (c)}
\\
\includegraphics[width=.73\columnwidth]{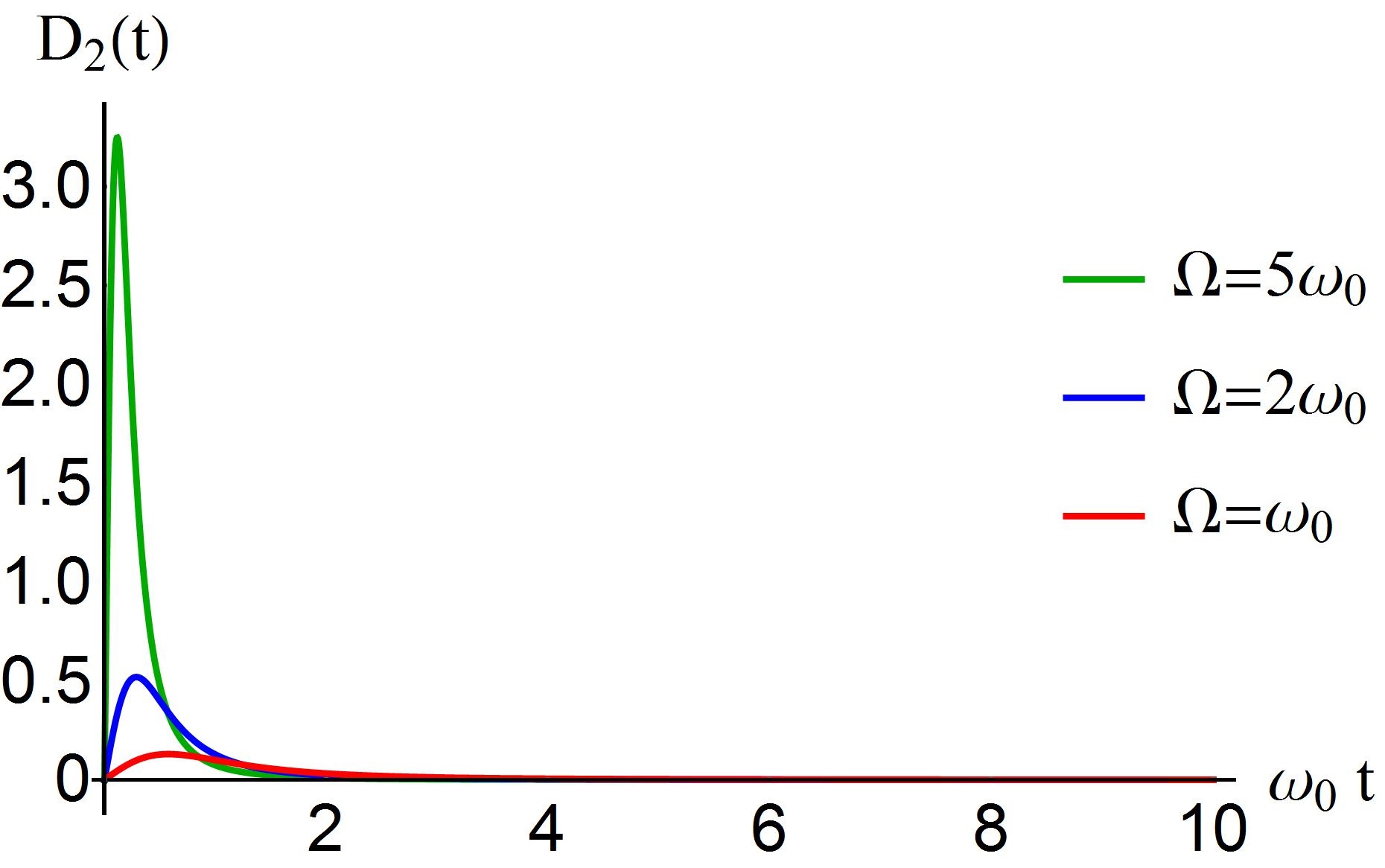}
\caption{(Color online) {\bf (a)} Time evolution of the noise kernel for different choices of the cutoff frequencies $\Omega$ and temperature $T_E$. {\bf (b)} Time evolution of the environmental correlation function, inferred from the width of the noise kernel, normalized by its maximum value (attained at time $t=0$) for $T_E=5$ \BVrev{and} different choices of cutoff frequency $\Omega$. {\bf (c)} Time evolution of the dissipation kernel for different choices of the cutoff frequencies $\Omega$.}
\label{fig3}
\end{figure}

In order to understand the behavior of the energy backflow shown in Fig. \ref{fig:ebf}, one has to consider in some detail the dependence on the relevant parameters $\Omega$ and $T_E$ of both the maximum and the correlation time of the noise and dissipation kernels $D_1(t)$ and $D_2(t)$, as given by \eqref{eq:D1D2}. 
These behaviors are shown in Figs. \ref{fig3} \textbf{(a)}, \textbf{(b)}, \textbf{(c)}, where the correlation time of noise kernel can be inferred from the width of the ratio $D_1(t)/D_1(0)$.
More precisely the observed vertical gradient in Fig. \ref{fig:ebf} can be traced back to the varying amplitude of the noise kernel, whose maximum increases with growing temperature [see Fig. \ref{fig3}\textbf{(a)}], where one has to compare the initial value of the solid lines with the one of the dashed lines relative to the same $\Omega$. 
Similarly, the observed horizontal gradient in Fig. \ref{fig:ebf} is mainly determined by the correlation time of the noise kernel, which decreases with growing cutoff frequencies; see Fig. \ref{fig3}\textbf{(b)}. For fixed temperature $T_E$, the correlation time of the noise kernel decreases for \GGrev{growing values of the} cutoff frequency, so that for very large \GGrev{$\Omega$} the bath has a very short correlation time, which, in turn, is known to lead to a semigroup dynamical regime. This last horizontal trend, however, is compensated in the low-temperature region by the opposite behavior of the amplitude of the dissipation kernel which increases with growing cutoff frequency; see Fig. \ref{fig3}\textbf{(c)}.

\begin{figure}[!ht]
\includegraphics[width=1.03\columnwidth]{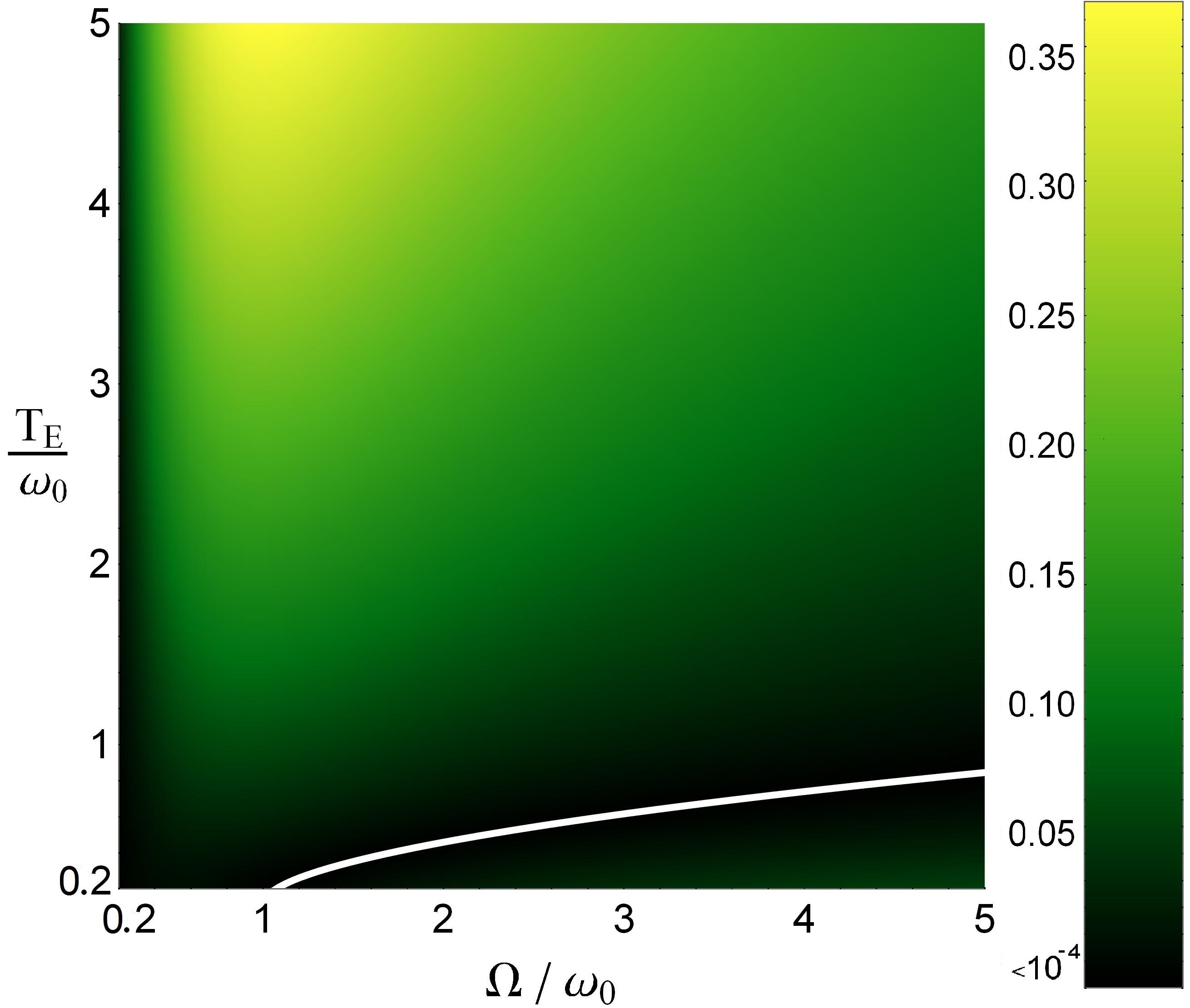}
\caption{(Color online)  Plot of the absolute value of $ \partial_{\omega}J_{eff}(\omega, \Omega,T_E)|_{\omega=\omega_0} $, for $\lambda=0.1$ and values of the cutoff frequency and environmental temperature $\left(\Omega/\omega_0,T_E/\omega_0\right)$ in the range $(0.2,5)\times(0.2,5)$. The black region of this plot, which has to be compared with the one in Fig. \ref{fig:ebf}, indicates those values of the parameters for which the resonance condition \eqref{resonancecondition} approximately holds, while the white curve denotes those for which \eqref{resonancecondition} strictly holds.}
\label{fig:dev}
\end{figure}

Finally, in order to explain the region of parameters where the energy backflow is suppressed (black region in Fig. \ref{fig:ebf}), we have to consider the behavior of the effective spectral density.
In particular $J_{eff}(\omega,\Omega,T_E)$ possesses one maximum $\omega_{max}$ with respect to its $\omega-$dependence, around which the dominant environmental modes are distributed.
Following the discussion in \cite{ClosBreuer}, which is recovered more thoroughly in Sec. \ref{sec:Markov}, if such maximum $\omega_{max}$, identified by the condition $\partial_{\omega}J_{eff} (\omega,\Omega,T_E) = 0$, is equal to the system's transition frequency $\omega_0$, then one has the resonance condition
\begin{equation}\label{resonancecondition}
\frac{\partial}{\partial\omega}J_{eff}(\omega,\Omega,T_E){}_{|\omega=\omega_0} = 0.
\end{equation}
Figure \ref{fig:dev} displays \BVrev{the absolute value of $ \partial_{\omega}J_{eff}(\omega, \Omega,T_E){}_{|\omega=\omega_0} $ for all the values $\left(\Omega/\omega_0,T_E/\omega_0\right)$ in the range $(0.2,5)\times(0.2,5)$, displayed on a colored scale}, showing the deviation from the resonance condition \eqref{resonancecondition}, denoted by the white curve in the plot.
It is immediate to see that energy backflow is almost suppressed (black region in Fig. \ref{fig:ebf}) whenever these deviations are small, i.e., when the resonance condition approximately holds.

Our analysis further provides a tool to identify the parameters region in which the energy backflow shows a maximum value. From Fig. \ref{fig:ebf} it is, in fact, evident that this condition is reached for high values of the temperature $T_E$ and for values of the cutoff \GGrev{frequency $\Omega$ around the system proper frequency $\omega_0$.}

\section{Correspondence between Energy Backflow and non-Markovianity}
\label{sec:Markov}

In the present section we discuss in detail the connection between our thermodynamic quantity, namely the energy (back-)flow between a reduced system and its environment, obtained in \BVrev{the} FCS formalism, and the concept of non-Markovianity. Among the different criteria and measures introduced so far to define and quantify non-Markovianity \cite{Wolf2008PRL,Breuer2009PRL,Breuer2012JPB,Rivas2010PRL,Lu2010PRA,
LuoPRA2012,Lorenzo2013PRA,Bylicka2013arxiv,Chruscinski2014PRL}, we concentrate our attention to the one introduced by Breuer, Laine, and Piilo \cite{Breuer2009PRL}, which has also been experimentally measured in all-optical settings \cite{Liu2011NAT,Tang2012EPL,Liu2013SCI}. 
The reason for this choice is the physical interpretation behind it, which we briefly recall. According to \cite{Breuer2009PRL}, any change in the distinguishability between two reduced states can be read in terms of an information flow between system and environment. Such distinguishability is quantified through the trace distance \cite{Nielsen2000}, which is a metric on the space of states induced by the trace norm and has the property to be a contraction under the action of completely positive and trace-preserving maps (i.e., physically implementable channels).
The evolution of the trace distance between two states of the reduced system coupled to the same environment but evolved from different initial conditions, which we denote with
\begin{equation}\label{trd1}
D(t, \rho_S^{1,2}) \equiv D(\rho^1_S(t), \rho^2_S(t)),
\end{equation}
describes the information exchange between the system and its environment. In particular, a decrease of \eqref{trd1} indicates a \BVrev{reduced} ability to discriminate between the two initial conditions $\rho_S^{1}(0)$ and $\rho^2_S(0)$, this in turn meaning that some information has flown out of the system towards the environment. Analogously, a temporary increase of the trace distance can be ascribed to a backflow of information from the environment to the system again.
Non-Markovian quantum dynamics are accordingly defined as those which show a non-monotonic behavior of the trace distance, i.e. such that there exist time intervals where
\begin{equation}\label{BLPsignature}
\sigma(t,\rho_S^{1,2}) = \frac{d}{dt} D(t,\rho_S^{1,2}) >0.
\end{equation}
Building on this definition, the non-Markovianity measure introduced in \cite{Breuer2009PRL} is just the sum of the trace-distance regrowths, conveniently maximized over all possible couples of initial states of the system:
\begin{equation}\label{eq:nblp}
\mathscr{N}= \max_{\rho_S^{1,2}(0)} \,\frac{1}{2}\int_0^{+\infty} dt\, \left(\left|\sigma(t,\rho^{1,2})\right|+\sigma(t,\rho^{1,2})\right).
\end{equation} 
This measure of non-Markovianity has already been calculated for the spin-boson model in \cite{ClosBreuer}, where it turned out to be a function of the temperature of the environment $T_E$ and of the cut-off frequency $\Omega$. We will comment in more detail about this in a short while.

Let us proceed to discuss the connection between the occurrence of non-Markovianity and of energy backflow, the latter being witnessed by the time behavior of $\theta(t)$ according to \eqref{condenb}.
First of all, we have already shown that in the Born-Markov approximation the energy flow per unit of time \eqref{theta} becomes a positive nonoscillating function; see Fig. \ref{fig1} \textbf{(b)} and Eqs. \eqref{thetaLT} and \eqref{rhoLT}. This physically corresponds to the case of a system which \BVrev{steadily loses energy to the environment with a positive rate}, i.e., a monotonic unidirectional flow of energy pointed towards the environment. Moreover, since such limiting dynamics corresponds to a quantum dynamical semigroup, i.e., the master equation reduces to GKSL form \cite{ClosThesis,GKSL}, the non-Markovianity measure vanishes. More strongly, semigroup dynamics actually represents a Markovian dynamics for every criterion so far introduced, \BVrev{making this particular result independent from the definition of non-Markovianity chosen.}
Apart from this limiting case the function $\theta(t)$ has been proven to oscillate in time for every value of the relevant parameters, namely cutoff frequency, as well as temperatures of the bath and of the system and coupling strength (always within weak-coupling regime that allows the second-order time-convolutionless expansion of the GME).
Such oscillations reflect the time behavior of the rate by which the system loses its energy in favor of the environment: If $\theta(t)$ remains positive, then the energy flow still remains unidirectionally pointed from the system towards the environment. If, however, for certain values of the parameters, $\theta(t)$ temporarily takes on negative values, then energy backflow occurs. 

As mentioned above, the non-Markovianity measure \eqref{eq:nblp} for the spin-boson model has already been evaluated for this model \cite{ClosBreuer}, where the calculations have been carried out for a slightly different choice of the spectral density $J(\omega)$. In \cite{ClosBreuer}, in fact, the authors employed a Lorentzian cutoff instead of an exponential one [see our Eq. \eqref{SpectralDensity} and their Eq. (19)]; we have reevaluated such measure $\mathscr{N}(\Omega,T_E)$ with the current spectral density and the result has turned out to be substantially unchanged, this confirming the suggestion that the information backflow does not significantly depend on the high-frequency part of the spectrum \cite{Addis}, at least in the case of a bosonic bath. Therefore, in order to avoid redundancy, we limit ourselves to recall here the most significant features of this measure, referring to Appendix \ref{app:NM} for the plot of $\mathscr{N}(\Omega,T_E)$.
First, for large values of the cutoff frequency \BVrev{$\Omega$ ($\simeq 10\omega_0$)} the spectral density can be approximated with $J(\omega) \simeq \lambda \omega $, this leading to a Markovian dynamics.
On the other hand, for decreasing values of the cutoff frequency $\Omega$ the amount of non-Markovianity, in general, increases, the only exception being represented by the region of parameters in which the resonance condition \eqref{resonancecondition} holds.
Such condition expresses the requirement of local flatness of the effective spectral density around the system's transition frequency, and describes a curve in the $\left(\Omega\,,\,T_E\right)$ plane called resonance curve along which a predominantly Markovian regime is expected and found \cite{ClosBreuer}.
In the case here considered of exponential cutoff, the resonance curve, which reads 
\begin{equation}\label{rescurve}
\Omega_{res}(T_E) = \frac{T_E}{\frac{T_E}{\omega_0}-\mathrm{Cosech}\left(\frac{\omega_0}{T_E}\right)},
\end{equation}
continues to retrace well the observed Markovian region at low temperatures of the bath (see Fig. \ref{fig:nM} in Appendix \ref{app:NM}).

A comparison between Figs. \ref{fig:ebf} and \ref{fig:nM} clearly shows that \GGrev{the amount of non-Markovianity of the dynamical map as measured by \eqref{eq:nblp} and the amount of energy backflow as quantified by \eqref{enbackflow}} are connected to each other.
First of all, in fact, for every value of the cutoff frequency $\Omega$, both quantities increase with increasing values of the temperature, this being related to the fact that in this model, when $T_E$ grows, the lower frequency part of the effective spectrum $J_{eff}(\omega,\Omega,T_E)$ is enhanced.
Moreover, both the non-Markovianity \GGrev{and the energy backflow measures} generally increase for decreasing values of the cutoff frequency. This is due to the fact that the correlation time of the noise kernel reduces for growing values of $\Omega$, so that the correlation function of the bath is almost $\delta$ correlated in time, which leads to a semigroup (and therefore Markovian) dynamics.
Finally, as already highlighted in Sec. \ref{sec:SB} A, both quantities are strongly related to the resonance condition \eqref{resonancecondition}. In particular, while the non-Markovianity measure \eqref{eq:nblp} vanishes only when \eqref{resonancecondition} holds strictly, the energy backflow is suppressed even when \eqref{resonancecondition} is approximately satisfied.
This result, together with the one discussed above in the Born-Markov regime, makes it possible to conclude that in a Markovian regime energy backflow is suppressed.
The opposite is, however, in general, not true; namely, the absence of energy backflow does not imply absence of information backflow, \Brev{thus preventing a one-to-one relation between these two concepts, as expected from both a mathematical and a conceptual point of view.}

The occurrence of energy backflow, in conclusion, \Brev{appears} as a stricter condition than non-Markovianity.
On the other hand, however, for values of the parameters $\Omega$ and $T_E$ for which the amount of non-Markovianity is significant, it becomes possible to measure a backflow of energy, as witnessed by the colored region in Fig. \ref{fig:ebf}.
\GGrev{We also stress that the relationship between the amount of energy backflow and non-Markovianity has to be intended at the level of the respective measures \eqref{enbackflow} and \eqref{eq:nblp}, which are properties of the dynamical map uniquely determined by the choice of the parameters $\lambda$, $\Omega$, and $T_E$.}
The connection we have found between non-Markovianity and energy backflow measures can \GGrev{finally} represent a powerful hint in relation to the practical usefulness of non-Markovianity: It is, in fact, clear from this result that a convenient engineering of the reservoir such to achieve non-Markovianity \cite{Haikka2011PRA,Verstraete} allows to have energy backflow and therefore to treat the environment as a potential quantum energy buffer.

\section{CONCLUSIONS}
\label{sec:Conclusions}

Using \BVrev{the FCS} formalism, we have studied the mean value of the energy exchange between a system of interest and its environment in the framework of the second-order time-convolutionless GME, introducing a suitable condition and quantifier for the occurrence of energy backflow from the environment back to the system.
We have then applied this construction to the spin-boson model, where we also showed how the first moment of the energy increment in the environment does not correspond only to the amount of energy lost by the system in the short and intermediate time scale, but has an additional oscillating contribution which reflects the quantum-mechanical feature of the interaction. Such deviation has been proven to vanish in the long-time limit, where the Born-Markov approximated solution faithfully describes the dynamics, given in terms of a semigroup.
Moreover, choosing an Ohmic spectral density with exponential cutoff to describe the distribution of bath modes and their interaction with the two-level system, we have studied the time behavior of the energy flow as a function of the many relevant parameters of the model, such as the environmental and effective system's temperatures, the cutoff frequency, and the coupling strength.
Results have shown that, for certain values of these parameters, the energy which has flown from the two-level system to the environment can effectively come back.
We point out that, while in this work we \BVrev{have} focused \GGrev{our attention} on the first cumulant of the exchanged energy, higher-order cumulants can also be considered using the FCS formalism \cite{Flindt}, making it possible to discuss, for example, bunching properties of bosons in the presence of energy backflow.
We have finally considered an important criterion of non-Markovianity, namely the one based on the time behavior of the trace distance
between two distinct initial states, and connected it, \Brev{relying on an analysis of the behavior of the effective spectral density at the system frequency, to the occurrence of energy backflow in the considered system}.
The comparative analysis has shown that non-Markovianity allows for the observation of energy backflow.
Our quantifier of energy backflow might also have interesting applications in the context of quantum thermodynamics of open quantum systems \cite{Nejad2015}, also in the connection with non-Markovianity, a topic recently attracting much attention \cite{Maniscalco2015}, as well as in the study of environment-induced entanglement \cite{ThirdReferee}.

\acknowledgments

This work was supported by JSPS Grant-in-Aid for Scientific Research (C) No. 15K05207 and JSPS Grant-in-Aid for Scientific Research (B) No. 25287098.
B. V. also acknowledges support by European Union (EU) through the Collaborative Projects QuProCS (Grant Agreement 641277), by the COST Action NO. MP1006 Fundamental Problems in Quantum Physics and by UniMI through the H2020 Transition Grant No. 14-6-3008000-623.

\appendix 

\section{Evaluation of the ground-state population $\rho_{00}(t)$}
\label{app:AppA}

In this appendix we explicitly give the differential equation for the ground-state population of the reduced system $\rho_{00}(t)$, as well as its formal solution, starting from the GME formalism of Eq. \eqref{matrixXichi}. 

As written in the main body of the paper, if we simply set the counting field parameter $\chi=0$, we obtain the usual master equation for the statistical operator of the reduced system $\rho(t)$. 
In particular, it is clear from \eqref{matrixXichi} that the dynamics of the coherences is decoupled from the one of the populations, and therefore  only the evolution of the latter determines the behavior of Eq. \eqref{theta}. It is therefore convenient to introduce the vector $\ket{\rho_d(t)} = \left(\rho_{00}(t),\rho_{11}(t)\right)^T$ and the $2\times2$ matrix $\Xi_d(t)$ obtained extracting the elements of $\Xi^{\chi=0}(t)$ relative to $\ket{\rho^{\chi}_d}$
\begin{equation}\label{Xid}
\Xi_d(t) = \begin{pmatrix}
a_+(t) & -a_-(t) \\ -a_+(t) & a_-(t)
\end{pmatrix},
\end{equation}
where $a_{\pm}(t) = -\int_0^t\,d\tau V_{\pm}(\tau)$ and where we have used the relation
\begin{equation}
W^{\chi=0}_{\pm}(\tau) = -V_{\mp}(\tau).
\end{equation}
The differential equation for the ground-state population $\rho_{00}(t)$ therefore reads
\begin{equation}\label{eq:rho00dot}
\frac{d}{dt}\rho_{00}(t) = (a_+(t)+a_-(t))\rho_{00}(t) - a_-(t),
\end{equation}
whose formal solution has the form
\begin{equation}\label{formalsolrho00}
\rho_{00}(t) = e^{\int_0^t\,d\tau a_{zz}(\tau)} \Bigl(\rho_{00}(0) -\int_0^t\,d\tau a_-(\tau) e^{-\int_0^{\tau}\,ds a_{zz}(s)}\Bigr),
\end{equation}
with 
\begin{equation}\label{azz}
a_{zz}(t) \equiv a_+(t) + a_-(t) =  -2 \int_0^t d\tau \, D_1(\tau) \cos(\omega_0\tau)
\end{equation}
being one of the right eigenvalues of $\Xi_d(t)$.
We finally notice that, for reasons of computation-time advantages, it is better to express the quantity $a_-(t)$ as
\begin{equation}\label{equivrel}
\quad a_-(t) = \frac{1}{2}\left(a_{zz}(t) + b_z(t)\right),
\end{equation}
where $a_{zz}(t)$, given by Eq. \eqref{azz}, and
\begin{equation}\label{definitions}
b_z(t) = -2 \int_0^t d\tau \, D_2(\tau) \sin(\omega_0\tau)
\end{equation}
have been usually employed in the treatment of the spin-boson model \cite{Breuer2002,ClosBreuer}. 
In fact, while both $ a_{zz}(t) $ and $ a_-(t) $ can only be numerically accessed, the quantity $b_z(t)$ can be analytically solved. This splitting of the nonhomogeneous term $a_-(t)$ \eqref{eq:rho00dot} into a numerical part $a_{zz}(t)$ and an analytic term $b_z(t)$ \eqref{equivrel} allows for shorter computation times.

The long-time limit approximation of the dynamics can be obtained by taking the limit for $t\to+\infty$ in Eq. \eqref{matrixXichi}, this corresponding to the Born-Markov approximation. 
The matrix \eqref{Xid} which governs the evolution of populations takes the form
\begin{equation}\label{XichidLT}
\bgreek{\Xi}^{\chi}_{d,LT} = \begin{pmatrix}
-\Gamma n(\omega_0) & \Gamma \left(1+n(\omega_0)\right) \\
\Gamma n(\omega_0) & -\Gamma \left(1+n(\omega_0)\right)
\end{pmatrix},
\end{equation}
where $\Gamma \equiv 2\pi J(\omega_0)$.
In order to arrive to this expression, the general relation
\begin{equation}\label{PSrelations}
\int_0^{+\infty}\,d\tau e^{i(\omega-\omega_0)\tau} = \pi\delta\left(\omega-\omega_0\right) + i\mathscr{P} \frac{1}{\omega-\omega_0}
\end{equation}
has been used \cite{UchiyamaPRE}.
As a consequence, the differential equation for the ground-state population \eqref{eq:rho00dot} becomes
\begin{equation}\label{eq:rho00dotLT}
\frac{d}{dt}\rho_{00,LT}(t) = -\Gamma \left(1+2n(\omega_0)\right)\rho_{00,LT}(t) + \Gamma\left(1+n(\omega_0)\right),
\end{equation}
whose solution reads
\begin{equation}\label{rhoLT}
\rho_{00,LT}(t) = \frac{1+n(\omega_0)}{1+2n(\omega_0)}+\rho_{00}(0) e^{-\Gamma\left(1+2n(\omega_0)\right)t}.
\end{equation}
Using \eqref{PSrelations} we can also finally compute the expression of the long-time limit version of the energy flux per unit of time $\theta_{LT}(t)$. In fact, since
\begin{equation}
w_{+,LT} = \left[\frac{\partial}{\partial (i\chi)} \int_0^{+\infty} \,d\tau W^{\chi}_+(\tau)\right]_{\chi=0} \!\!\!\!= -\omega_0\Gamma\left(1+n(\omega_0)\right)
\end{equation}
and
\begin{equation}
w_{-,LT} = \left[\frac{\partial}{\partial (i\chi)} \int_0^{+\infty} \,d\tau W^{\chi}_-(\tau)\right]_{\chi=0} = \omega_0\Gamma n(\omega_0),
\end{equation}
the expression for $\theta_{LT}(t)$ becomes
\begin{equation}\label{thetaLT}
\theta_{LT}(t) = 
\omega_0\frac{d}{dt}\rho_{00,LT}(t).
\end{equation}
The integral form of this expression gives the result
\begin{equation}
\langle\Delta q\rangle_{t,LT} = \omega_0\left(\rho_{00,LT}(t)-\rho_{00}(0)\right).
\end{equation}

\section{Proof of Equation (30)}
\label{app:AppB}

In this appendix we \GGrev{show} the detailed calculations required to arrive at expression \eqref{relationimp} for the energy flow per unit of time $\theta(t)$ starting from \eqref{theta}.
First, since the simple identity
\begin{equation}\label{derivativeidentity}
\left.\frac{\partial \Phi(\pm \tau-\chi)}{\partial (i\chi))}\right|_{\chi=0} = \pm i \frac{\partial \Phi(\pm\tau)}{\partial \tau}
\end{equation} 
holds, it becomes possible to reexpress both the terms $w_+(t)-w_-(t)$ and $w_+(t)$ that appear in \eqref{theta} in an equivalent form. In particular, one gets
\begin{align}
&w_+(t)\!-\!w_-(t)\!=\! 2\int_0^t\,d\tau\,\left(\partial_{\tau}D_1(\tau)\right)\sin(\omega_0 t), \notag\\
&w_+(t)\!=\!\int_0^t\!\!d\tau\left(\partial_{\tau}D_1(\tau)\right)\sin(\omega_0 t)\!-\!\!\!\int_0^t\!\!d\tau\left(\partial_{\tau}D_2(\tau)\right)\cos(\omega_0 t).
\end{align}
An integration by parts of the quantities above, using $ D_1(0)\sin(0) = 0 $ and $D_2(0)=0$ and Eqs. \eqref{eq:rho00dot}, \eqref{azz}, \eqref{equivrel}, and \eqref{definitions}, then gives
\begin{align}
& w_+(t) - w_-(t)= 2 D_1(t) \sin(\omega_0 t) + \omega_0 a_{zz}(t),\notag\\
&w_+(t) = D_1(t) \sin(\omega_0 t) - D_2(t) \cos(\omega_0 t) + \omega_0 a_-(t),
\end{align}
from which Eq. \eqref{relationimp} immediately follows.

\section{Plot of the non-Markovianity measure $\mathscr{N}(\Omega,T_E)$} \label{app:NM}

\begin{figure}[!ht]
\includegraphics[width=1.04\columnwidth]{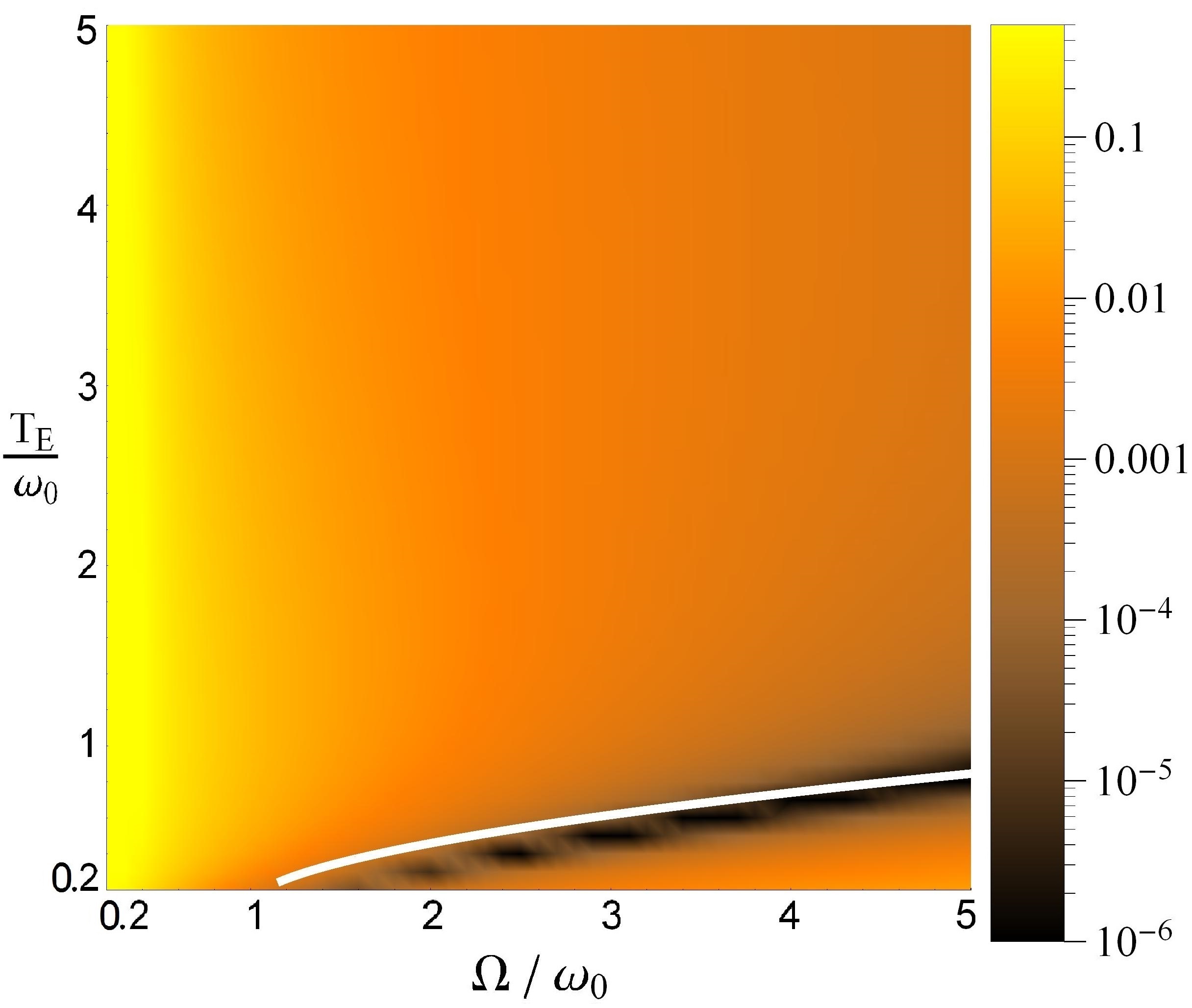}
\caption{(Color online)  Plot of the non-Markovianity measure $\mathscr{N}(\Omega,T_E)$ up to integration time $t=100\omega_0^{-1}$, for $\lambda=0.1$ and for values of the parameters $\left( \Omega/\omega_0, T_E/\omega_0\right)$ in the range $(0.2 , 5) \times (0.2 , 5)$. The white line corresponds to the resonance curve \eqref{rescurve}, which provides an approximate estimate of the region of Markovianity of the dynamics}.
\label{fig:nM}
\end{figure}

We give in this appendix the plot of the non-Markovianity measure for the spin-boson model described by the Hamiltonian \eqref{eq:Hamiltonian} and calculated for a spectral density of the form \eqref{SpectralDensity}. Figure \ref{fig:nM} shows $\mathscr{N}(\Omega,T_E)$ for $\lambda=0.1$ and for values of the parameters $\left( \Omega/\omega_0, T_E/\omega_0\right)$ in the range $(0.2 , 5) \times (0.2 , 5)$, chosen as in Fig. \ref{fig:ebf}. The couple of initial states $\rho^1(0)$ and $\rho^2(0)$ used is the one that maximizes \eqref{eq:nblp} in accordance with \cite{ClosBreuer}, namely those with Bloch vectors $v^1(0) = (0,1,0)^T$ and  $v^2(0)= (0,-1,0)^T$. Finally, the upper limit in the integral \eqref{eq:nblp} has been chosen equal to $t=100\omega_0^{-1}$. A comparison with Fig. 3 of \cite{ClosBreuer} shows that the change in the high-frequency part of the spectral density does not affect significantly the non-Markovianity measure.

The white line in Fig. \ref{fig:nM} finally shows the resonance curve \eqref{rescurve} around which the dynamics result to be Markovian.

\end{document}